\def\cm2{cm$^{-2}$}

\def\c2{C~{\sc ii}}
\def\c4{C~{\sc iv}}
\def\fe2{Fe~{\sc ii}}
\def\fe3{Fe~{\sc iii}}
\def\mg1{Mg~{\sc i}}
\def\mg2{Mg~{\sc ii}}
\def\si2{Si~{\sc ii}}
\def\si4{Si~{\sc iv}}
\def\al2{Al~{\sc ii}}
\def\al3{Al~{\sc iii}}
\def\o1{O~{\sc i}}
\def\n1{N~{\sc i}}
\def\h1{H~{\sc i}}

\def\approxlt{\mathrel{\spose{\lower 3pt\hbox{$\sim$}}
        \raise 2.0pt\hbox{$<$}}}
\def\approxgt{\mathrel{\spose{\lower 3pt\hbox{$\sim$}}
        \raise 2.0pt\hbox{$>$}}}

\def\rr{\hbox{RR Lyrae~}}

\documentclass[article]{gwp80}
\usepackage{graphicx}
\usepackage{amssymb}
\tabletypesize{\normalsize}  

\shortauthors{Bono et al.}
\shorttitle{RR Lyrae the Stellar Beacons of the Galactic Structure}

\begin{document}
\large    
\pagenumbering{arabic}
\setcounter{page}{1}

\title{RR Lyrae the Stellar Beacons of the Galactic \\ \\ Structure}

%
%
\author{{\noindent Giuseppe Bono{$^{\rm 1,2}$}, Massimo Dall'Ora{$^{\rm 3}$}, Filippina Caputo{$^{\rm 2}$},
Giuseppina Coppola{$^{\rm 3}$}, Katia Genovali{$^{\rm 1}$}, Marcella Marconi{$^{\rm 3}$}, Anna Marina
Piersimoni{$^{\rm 4}$}, Robert F. Stellingwerf{$^{\rm 5}$} \\
\\
{\it (1) Universit\'a di Roma Tor Vergata, Roma, Italy;\\
     (2) INAF-Osservatorio Astronomico di Roma, Italy;\\
     (3) INAF--Oservatorio Astronomico di Capodimonte, Napoli, Italy;\\
     (4) INAF--Oservatorio Astronomico di Collurania, Teramo, Italy;\\
     (5) Stellingwerf Consulting, 11033 Mathis Mtn Rd SE, 35803 Huntsville, AL USA\\
}
}
}

%
%
\email{(1) bono@roma2.infn.it (2) massimo.dallora@na.astro.it, (3) filippina.caputo@oa-roma.inaf.it,
(4) coppola@na.astro.it, (5) katiagenovali@hotmail.com, (6) marcella.marconi@oacn.inaf.it,
(7) piersimoni@oa-teramo.inaf.it, (8) rfs@swcp.com
}

\begin{abstract}
We present some recent findings concerning the use of RR Lyrae as distance indicators 
and stellar tracers. We outline pros and cons of field and cluster RR Lyrae stars and 
discuss recent theoretical findings concerning the use of the Bailey (amplitude vs 
pulsation period) diagram to constrain the possible occurrence of Helium enhanced 
RR Lyrae stars. Nonlinear, convective RR Lyrae models indicate that the pulsation 
properties of RR Lyrae stars are minimally affected by the helium content. The main 
difference between canonical and He enhanced models is due to the increase 
in luminosity predicted by evolutionary models. Moreover, we focus our attention on 
the near-infrared Period-Luminosity (PL) relation of RR Lyrae and summarize 
observational evidence concerning the slope of the K-band PL relation in a 
few globulars (M92, Reticulum, M5, $\omega$ Cen) covering a range in metallicity of 
$\sim$1 dex. Current findings suggest that the slope has a mild dependence on the metal 
content when moving from the metal-poor to the metal-intermediate regime. 
Finally, we also discuss the use of RR Lyrae stars either to estimate (helium 
indicator: A-parameter) or to measure (absorption and emission lines) the helium 
content.   
\end{abstract}
\vbox{\vskip18pt}

\section{Introduction}

Even after a cursory reading of the papers published in the seminal Vatican conference 
on stellar populations it becomes clear the key role that RR Lyrae played in the development 
of what we now call stellar populations and in the shaping of Galactic structure. 
Their relevance  is also supported by the first sentence of one the papers given by 
W. Baade at that conference: 

{\em Variable stars, in particular the cepheids and the cluster-type variables, have 
becoming increasingly important in the exploration of our own and other galaxies.}     

The relevance of RR Lyrae stars in stellar astrophysics is further supported by 
the fact that RR Lyrae are the most popular old, low-mass distance indicators.  
In early times it was assumed that their mean luminosity was constant, but soon after 
the spectroscopic investigations showed that they do obey a visual 
magnitude metallicity relation. Therefore, the knowledge of their metallicity 
was fundamental for accurate estimates of their individual distances.   
However, high-resolution spectroscopy in single slit mode was prohibitive 
for large numbers of RR Lyrae stars. The $\Delta$S method invented by 
George \citep{Preston59} opened a new path in the extensive use of field RR Lyrae 
as stellar tracers of the Galactic halo \citep{Suntzeff94} and of 
the bulge \citep{wal91,blan97}. 

RR Lyrae are also fundamental laboratories to constrain the physical mechanisms 
that drive their pulsation instability. After the pioneering investigations by 
\citet{Christy66} based on nonlinear radiative models, the modelling of RR Lyrae 
pulsation behavior was at the cross-roads of several theoretical and observational 
investigations. Once again the papers by \citet{Preston64} and \citet{Preston65} have been 
a benchmark for those interested in understanding the impact that the different 
physical mechanisms (convection) and input physics (opacity, equation of state) 
have on light and radial velocity curves. 

The structure of this paper is the following. In \S2 we briefly discuss pros and cons 
of cluster and field RR Lyrae. The pulsation framework developed by our group during 
the last few years is outlined in \S3, together with the added values and the drawbacks 
of the Bailey Diagram (amplitude versus period). In \S5 we discuss the current 
observational scenario of the K-band Period-Luminosity relation of cluster RR Lyrae.   
In \S6 we discuss the diagnostics currently adopted to estimate the primordial 
helium content, in particular we focus our attention on RR Lyrae stars. In the last 
section we briefly mention the possible paths that the next generation of observing 
facilities will open during the next few years.

\section{Pros and cons of field and cluster RR Lyrae stars}  

Field and cluster RR Lyrae bring forward several distinctive advantages worth 
discussing in detail:

{\em i)}--Distance-- Cluster RR Lyrae provide accurate distance estimates of 
Galactic Globular Clusters (GGCs), 
and in turn more accurate estimates of their absolute age. Unfortunately, 
they are not ubiquitous, i.e. their occurrence in GGCs does depend on the 
morphology of the horizontal branch. They are not present in GGCs that have 
either a very red or a very blue HB morphology. This means that they are 
affected by the so-called second parameter problem (e.g., Kunder et al. 2011).

{\em ii)}--Ensemble properties-- Cluster and field RR Lyrae stars do show different 
period distributions, i.e. they are affected by the so-called Oosterhoff dichotomy 
\citep{oos39,Bono95a,smith95,cate09}. 
However, RR Lyrae in MC globulars are Oosterhoff intermediate \citep{Bono94a}. 
The same outcome applies to RR Lyrae in different dwarf spheroidal galaxies in 
the Local Group (e.g. Carina, \citealt{Dallora03}).  The fraction of first overtone 
(FO, RRc) and fundamental (F, RRab) variables is also affected by the HB morphology and 
changes when moving from GGCs to dwarf galaxies to the Galactic field \citep{Petroni03}. 
This indicates that pulsation properties of RR Lyrae 
do depend on the chemical and dynamical evolution of their stellar environment.  

{\em iii)}--Evolutionary status-- Empirical evidence and theoretical predictions 
indicate that the range in visual magnitude between the Zero-Age-Horizontal-Branch 
(ZAHB) and the end of central helium burning is correlated with the metallicity
\citep{Sandage90,Bono95b}. Together with this intrinsic evolutionary 
effect, we are also facing the problem that we still lack a firm photometric 
diagnostic to constrain the evolutionary status of individual RR Lyrae stars.     
This drawback applies to both cluster and field RR Lyrae and causes a series 
of problems not only in the distance estimate, but also in the estimate of the 
observed ZAHB luminosity level, and in turn on the parameters correlated with 
this evolutionary feature, namely the R parameter (\citealt{Sandquist00}; 
Troisi et al.  2011, in preparation), the  $\Delta$$V_{bump}^{HB}$ 
\citep{Dicecco10} and the relative ages of GGCs \citep{Marin09}.    
A possible solution to overcome this problem is to estimate the individual distances 
using the NIR PL relations, since they are minimally affected by evolutionary 
effects and by a possible spread in mass inside the instability strip \citep{Bono01,Bono03}. 
On the basis of the true distance modulus and of their individual reddenings,  
the same objects can be located in the absolute CMD ($M_V$ vs $(B-V)_0$) to constrain 
their evolutionary status (Bono et al. 2011, in preparation).        

{\em iv)}--Evolutionary/Pulsation connection-- Cluster variables are also relevant  
to constrain the input physics adopted in pulsation and evolutionary codes. Observables 
predicted by pulsation models (periods, modal stability, pulsation amplitudes) and by 
evolutionary models (lifetimes, mass-luminosity relation, effective temperature) do 
depend on the same intrinsic parameters (stellar mass, chemical composition). The 
comparison between theory and observations using the same objects (RR Lyrae),  
seen as variables and as HB stars, provide a fundamental sanity check for both 
the micro (opacity, equation of state, cross sections) and the macro 
(gravitational settling, mixing, mass-loss) physics. This is a fundamental 
stepping stone to improve the plausibility of the physical assumptions adopted 
in pulsation and evolutionary models, since the former rely on the mass-luminosity 
relation predicted by evolutionary models. This is the path that has been 
followed to address several open problems (Oosterhoff dichotomy, Sandage period 
effect, hysteresis mechanism, topology of the instability strip) that are at the
cross-roads between theory and observations.    

{\em iv)}--Statistics and progenitors-- The number of GGCs that host several tens 
of RR Lyrae stars is quite limited \citep{Clement01}. The new photometric surveys
are disclosing several thousands of field RR Lyrae. This means that robust comparisons 
can only be based on field stars. However, for cluster variables we have accurate 
knowledge of the absolute age and chemical composition of the progenitors. For field 
RR Lyrae we can either estimate (photometric indices) or measure (spectroscopy) the 
chemical composition, but we can barely constrain their absolute age, and in turn 
the mass of the progenitor.   

{\em v)}--Helium abundance-- More than 30 years ago it was suggested that the 
mass-luminosity ratio of RR Lyrae, the so-called A-parameter \citep{Caputo83} 
can be adopted, via the pulsation relation \citep{Vanalbada73,cap98}, 
to estimate the helium content of individual variables \citep{Sandquist00}. 
The application of this diagnostic to cluster RR Lyrae is very powerful, since the 
precision is only limited by the size of the sample. The two main drawbacks of 
this approach is that the He estimates are affected by the precision of the 
color-temperature relation and by the evolutionary status of individual objects. 
A breath of fresh air on this delicate topic arrived with the discovery by 
\citet{Preston09} of both He~I and He~II emission and absorption 
lines in field RR Lyrae stars. The lines show up during the rising branch of 
RR Lyrae stars and appear to be the consequence of the shock propagation 
\citep{Bono94b,Chadid96}.

\section{Theoretical Framework}

The content of \S2 indicates that RR Lyrae stars are the cross-roads of several 
open astrophysical problems. Several of these problems call for new theoretical 
and observational insights. In particular, the theoretical approach does require 
a theoretical framework dealing with both evolutionary and pulsation properties
of radial variables. Our group undertook this project almost twenty years ago 
\citep{cap08}. In the following, we briefly mention recent findings concerning 
the predicted Bailey diagram and the K-band Period-luminosity relation.      

We constructed new sets of \rr pulsation models by using the hydrodynamical
code developed by \citet{Stellingwerf82} and updated by \citet{Bono94b} 
[see also \citealt{Smolec10} for a similar approach], \citet{Bono99}. 
The physical assumptions adopted to compute these models will be described in 
Marconi et al. (2011, in preparation). We adopted the OPAL radiative opacities 
released in 2005 by \citet{Iglesias96}\footnote{http://opalopacity.llnl.gov/} 
and the molecular opacities by \citet{Alexander94}.
To properly constrain the pulsation properties of \rr stars, we typically cover 
a wide range in metal abundances (scaled-solar, $\alpha$-enhanced). Moreover, 
to constrain the dependence of the pulsation properties of \rr stars on the 
He abundance we also adopt different values of this crucial parameter. 
For each fixed chemical composition the stellar mass of \rr stars
was fixed by using the evolutionary prescription for $\alpha$-enhanced structures provided by
\citet{Pietrinferni06} and available on the BaSTI database\footnote{http://albione.oa-teramo.inaf.it/}.
Note that for each fixed metal content the mass of the He-enhanced models was estimated
assuming the same cluster age (13 Gyr).
Together with the luminosity predicted by evolutionary models we often adopted a brighter
luminosity level to account for the possible occurrence of evolved
\rr stars. The reader interested in a more detailed discussion concerning evolutionary 
and pulsation ingredients is referred to \citet[][and references therein]{Marconi11}.

\subsection{The luminosity amplitude vs period (Bailey) diagram}

The period distribution and the Bailey diagram (luminosity amplitude vs pulsation period) 
are very robust observables, since they are independent of distance and reddening corrections. 
After the seminal discovery of this diagram \citep{oos39}, 
\citet{Preston59} found that objects with different  
metal abundances display different trends. The use of the pulsation amplitude as a proxy 
of the metal content is still lively debated both from the observational 
(\citealt{Kinemuchi06,Kunder09, Kunder11a}) and the theoretical \citep{Bono07,Fiorentino10} 
point of view. The reader interested in a detailed discussion concerning the development and 
the use of the Bailey diagram is referred to the detailed paper by Smith et al. 
(these proceedings). In the following, we focus our attention on the use of the Bailey 
diagram to constrain the helium content of RR Lyrae stars.    

The bolometric light curves predicted by nonlinear, convective, pulsation models 
are typically transformed into the observational plane using the bolometric
corrections and the Color-Temperature transformations provided by Castelli et al. (1997a,b).
Figure~1 shows predicted B-band amplitudes for fundamental (top) and first overtone (bottom) 
RR Lyrae models computed at fixed metal content (Z=0.001) and primordial helium contents 
ranging from Y=0.24 to Y=0.38 (see labeled values). To constrain the dependence of 
the luminosity amplitude on evolutionary effects the models --for each fixed chemical 
composition-- were constructed using two different luminosity levels. 

\begin{figure}
\centering
\includegraphics[width=9.0cm]{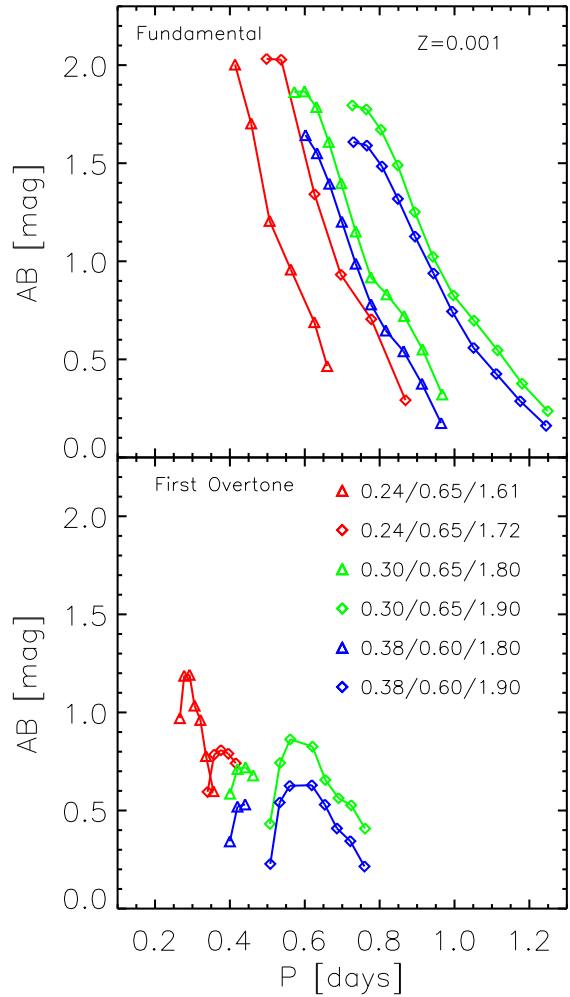}
\vspace*{1.55truecm}
\caption{Top -- Predicted B-band amplitudes versus period for fundamental RR Lyrae 
computed at fixed
metallicity
(Z=0.001). Different symbols 
display different luminosity levels, while  different colors show models 
constructed assuming different helium abundances and/or stellar 
masses. From left to right the labelled numbers show helium content 
(mass fraction), the stellar mass and the logarithmic luminosity 
(solar units).   
Bottom -- Same as the top, but for first overtone pulsators. 
}
\label{bailey}
\end{figure}

Data plotted in this figure indicate that the He content marginally affects the pulsation 
behavior of RR Lyrae. The top panel shows that an increase of 30\% in He content 
(0.24 vs 0.30) causes a systematic shift in the pulsation period. Thus, suggesting that 
the difference in the helium enhanced models is mainly due to the increase in the 
luminosity level. The different sets of models disclose that structures with a difference 
in helium content of $\sim$50\% might have, at fixed period, similar pulsation amplitudes. 
The amplitudes of the He enhanced models --Y=0.30, Y=0.38-- display minimal changes, since 
the latter group was constructed assuming the same luminosity levels and slightly smaller 
stellar masses (0.60 vs 0.65 $M_\odot$).   
The first the overtone pulsators (bottom panel) show similar trends concerning the He 
dependence.

This circumstantial evidence indicates that the Bailey diagram is not a good diagnostic 
to constrain the helium content of cluster and field RR Lyrae. Moreover, in dealing with 
luminosity amplitudes we need to keep in mind two relevant limits.
{\em i)}-- The theoretical Bailey diagram is affected by uncertainties on the mixing length
parameter \citep[see e.g.][]{Marconi03}. A decrease in the convective efficiency causes larger
amplitudes, but the pulsation periods are minimally affected. This means that the mixing 
length affects the slopes of the predicted amplitude--period relations, but the systematic 
drift as a function of the He content is not affected (Marconi et al. 2011, in preparation).  
{\em ii)}-- Recent space (COROT, \citealt{Chadid10}) and ground-based \citep{Kunder10}
observations indicate that the fraction of \rr stars affected by the Blazhko phenomenon 
is higher than previously estimated ($\approx$50\%, \citet{Benko10}.

\section{The NIR Period-luminosity relation of RR Lyrae stars}

RR Lyrae variables are relatively bright and have been detected in several Local Group galaxies 
(e.g.~\citealt{Dallora03,Dallora06,Pietrynzski08,Greco09,Fiorentino10,Yang10}) and 
can be easily identified from their characteristic light curves. 
The most popular methods to estimate their distances is to use either the visual 
magnitude--metallicity relation or the near-infrared (NIR) Period-Luminosity (PL) relation 
(e.g.~\citealt{Bono03b,Cacciari03}) or parallaxes and proper motions \citep{Feast08}.
The reader interested in independent approaches based on RR Lyrae to estimate stellar 
distances is referred to the thorough investigations by, e.g.,~\citet{Marconi03,Dicriscienzo04,Feast08}.

The visual magnitude--metallicity relation appears to be hampered by several uncertainties 
affecting both the zero-point and the slope~\citep{Bono03}. On the other hand, 
the NIR PL relation seems very promising, since it shows several relevant advantages. 
\citet{Longmore86} demonstrated, on an empirical basis, that RR Lyrae do obey 
to a well defined $K$-band PL relation.  The reason why the PL relation shows up in the 
NIR bands is due to the fact that the BC in the NIR bands, in contrast with the optical bands, 
steadily decreases when moving from the hot (blue) to the cool (red) edge of the RR Lyrae 
instability strip. This means that they become brighter as they become redder. 
The pulsation periods --at fixed stellar mass and luminosity-- become longer, since redder 
RR Lyrae have larger radii. The consequence of this intrinsic property is that periods and 
magnitudes are strongly correlated when moving toward longer wavelengths. 

Theoretical and empirical evidence indicates that the NIR PL relations of RR Lyrae
are robust methods to determine stellar distances.\\
{\em i)}-- The NIR PL relations are minimally affected by evolutionary effects 
inside the RR Lyrae  instability strip. The same outcome applies for the typical 
spread in mass inside the RR Lyrae instability strip~\citep{Bono01, Bono03}. 
Therefore, RR Lyrae distances based on the NIR PL relations are minimally 
affected by systematics introduced by their evolutionary status.
{\em ii)}-- Theory and observations indicate that fundamental (F) and 
first overtone (FO) RR Lyrae do obey independent NIR PL relations.
{\em iii)}-- Theory and observations indicate that the NIR PL relations are 
linear over the entire period range covered by F and FO pulsators.

The NIR PL relations also have three observational advantages. \\
{\em i)}-- The NIR magnitudes are minimally affected by uncertainties on 
the reddening. 
{\em ii)}-- The NIR amplitudes are at least a factor of 2-3 smaller than 
in the optical bands. Therefore, accurate estimates of the mean 
NIR magnitudes can be obtained with a modest number of observations. Moreover, 
empirical K-band light curve templates~\citep{Jones96} can be adopted to 
further improve the accuracy of the mean magnitudes. 
{\em iii)}-- Thanks to 2MASS, accurate samples of local NIR standard stars 
are available across the sky. This means that both relative and absolute NIR 
photometric calibrations can be easily accomplished.

The use of the NIR PL relations is also affected by two limits. \\
{\em i)}-- Empirical estimates of the slope of NIR PL relations show a significant 
scatter from cluster to cluster. They range from $\sim -1.7$ 
(IC$4499$, Sollima et al. 2006) 
to $\sim -2.9$ 
(M$55$, Sollima et al. 2006) and it is not clear whether the  
difference is intrinsic or caused by possible observational biases.
{\em ii)}-- Current predictions indicate that the intrinsic spread 
of the NIR PL relations decreases  by taking into account either the metallicity 
or the HB-type of the Horizontal Branch (HB) 
\citep{Bono03,Cassisi04,Catelan04,Delprincipe06}. 
Therefore accurate distance estimates of field RR Lyrae do require an estimate of 
the metallicity. However, no general consensus has been reached yet concerning the value 
of the coefficient of the metallicity term in the NIR Period-Luminosity-Metallicity 
(PLZ) relations.  The current estimates for the $K$-band range from 
$0.08$~\citep{Sollima06} to $0.23$~\citep{Bono03} mag/dex by using cluster and 
field RR Lyrae, respectively.

\begin{figure}
\centering
\includegraphics[width=8cm]{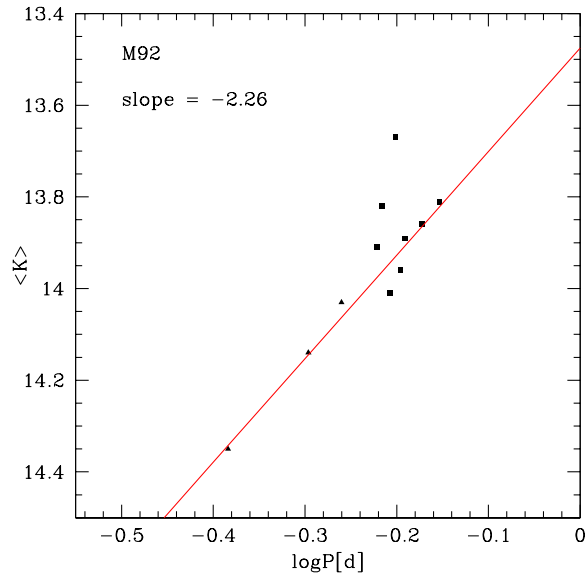}
\vskip0pt
\caption{K-band Period-Luminosity relation of RR Lyrae in the Galctic globular cluster M92, 
based on data collected by \citet{Delprincipe05}.  
The squares display fundamental (RRab) variables, while the triangles the first overtones (RRc). 
The latter were fundamentalized ($\log P_F$= $\log P_{FO}$ + 0.127). The red line shows the 
linear fit of the PLK relation and the slope is labeled.  
}
\label{M92}
\end{figure}

To address these problems our group undertook a long-term project aimed at providing 
homogeneous and accurate NIR photometry for several GCs hosting a good sample of 
RR Lyrae and covering a wide range of metal abundances. In the following, we briefly 
discuss the slope of the K-band\footnote{The intensity weighted mean magnitudes 
of the RR Lyrae discussed in this section were calibrated to the 2MASS  NIR photometric 
system using either local standards or the transformations provided by \citet{carp01}.} 
PL relation when moving from metal-poor to metal-intermediate GCs. 

\begin{figure}
\centering
\includegraphics[width=8cm]{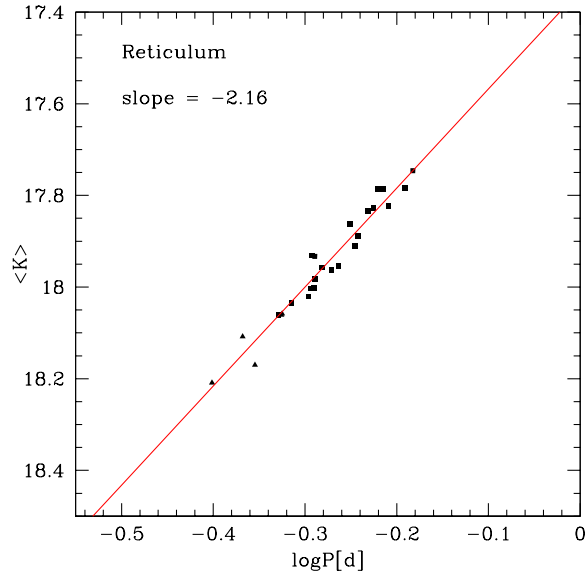}
\vskip0pt
\caption{Same as Fig.~2, but for RR Lyrae stars in the metal-poor LMC globular cluster Reticulum  
based on data collected by \citet{Dallora04}.  
}
\label{Reti}
\end{figure}

{\bf M92)}-- This is the most metal-poor cluster in our sample \citep{Delprincipe05}. 
According to the metallicity scale by \citet{Kraft03} based on FeII lines, the 
iron content is [Fe/H]=$-$2.38$\pm$0.07.  We observed eight fundamental and three first overtones. 
The latter were fundamentalized and we found a slope of $-$2.26$\pm$0.20, where the error 
only accounts for the uncertainty on the linear fit to the data. By using the K-band 
PL relation provided by \citet{Cassisi04} we found a true distance modulus  of $\mu$=14.62$\pm$0.04 mag,
that agrees quite well with similar estimates available in the literature 
\citet[][and references therein]{Dicecco10}.

{\bf Reticulum)}-- This is an LMC metal-poor cluster \citep{Dallora04}. 
According to the metallicity scale by \citet{Suntzeff92} based on Calcium triplet
lines, the iron content is [Fe/H]=$-$1.71$\pm$0.1. We observed 21 fundamental and 
five first overtones. 
The latter were fundamentalized and we found a slope of $-$2.16$\pm$0.09, where the error  
only accounts for the uncertainty on the linear fit to the data. By using the K-band 
PL relation provided by \citet{Bono03} we found a true distance modulus to Reticulum 
of $\mu$=18.523$\pm$0.005 mag, that agrees quite well with similar LMC distances 
available in the literature.

\begin{figure}
\centering
\includegraphics[width=8cm]{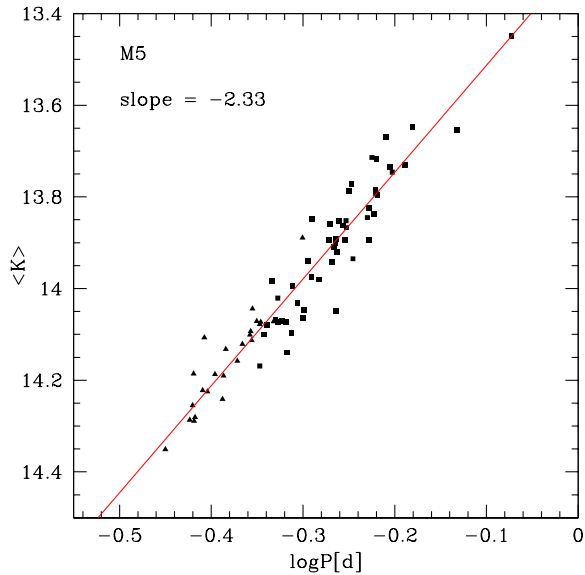}
\vskip0pt
\caption{Same as Fig.~2, but for RR Lyrae stars in the metal-intermediate globular cluster M5 
based on data collected by \citet{Coppola11}.  
}
\label{M5}
\end{figure}

{\bf M5)}-- This is the metal-intermediate ([Fe/H]=$-$1.26$\pm$0.06) cluster in our 
sample with the richest sample of first overtone (52) and fundamental (24) pulsators \citep{Coppola11}. 
By using the entire sample we found a slope of $-$2.33$\pm$0.08. We also found a true 
distance modulus to M5 of $\mu$=14.44$\pm$0.02 mag,  that agrees quite well with 
distances based on different distance indicators. The good agreement also applies 
to the kinematic distance. This seems an important finding, since this geometrical 
methods provides distance that are systematically larger than the distances based on
different distance indicators \citep{Bono08}.   

\begin{figure}
\centering
\includegraphics[width=8cm]{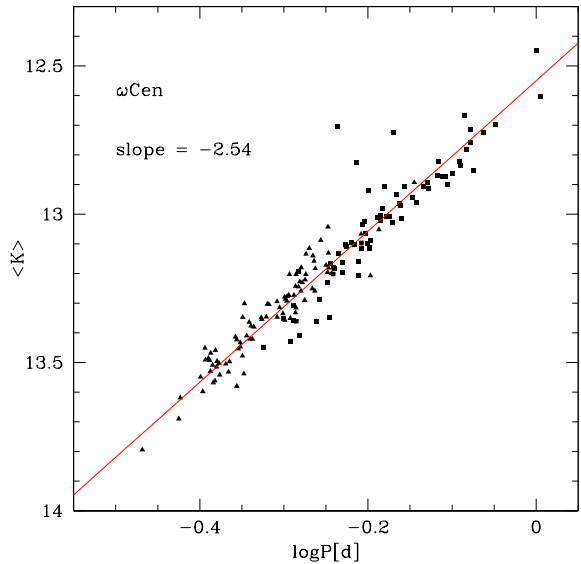}
\vskip0pt
\caption{Same as Fig.~2, but for RR Lyrae stars in the giant globular cluster $\omega$ Centauri 
based on data collected by \citet{Delprincipe06}.  
}
\label{omcen}
\end{figure}

{\bf $\omega$ Cen)}-- Finally, we also estimated the distance to $\omega$ Cen and observed 
93 first overtones and 104 fundamentals \citep{Delprincipe06}. The nature of this massive 
stellar system is not well established yet \citep{Bekki03}. However, there is general 
consensus that the stellar content of this system shows a metallicity distribution with 
multiple peaks \citep{Calamida09}. The same outcome applies to the heavy element 
abundances (\citealt{Johnson09, Pancino02}, and references therein).     
To test the dependence of the slope on the iron abundance we split the sample into 
a metal-poor and a less metal-poor subsample. We found that the slopes, within the errors, are very similar. 
Therefore, we decided to perform a linear fit over the entire sample and we found a slope 
of $-$2.54$\pm$0.09. By using the calibration of the K-band PL relation provided by 
\citet{Cassisi04} we found a true distance modulus to $\omega$ Cen 
of $\mu$=13.77$\pm$0.07 mag, that agrees quite well with distance estimates available 
in the literature \citep{Bono08}.     

The iron content of the GGCs we have already investigated, neglecting $\omega$ Cen, 
ranges from $-$2.38 to $-$1.26 dex. The slope of the K-band PL relation should change from 
$\sim$$-$2.1 \citep{Bono03} to $\sim$$-$2.4 \citep{Delprincipe06}. The former slope is based 
on models that include the metallicity term, while the latter include the HB type.  
Current evidence indicates that the slope of the K-band PL relation is marginally affected 
by iron abundance, thus supporting the results by by theory and by \citet{Sollima06}. 
However, more accurate data both in the metal-poor and in the metal-rich regime are 
required before we can reach firm conclusions concerning the metallicity dependence 
of both the slopes and the zero-points of the NIR PL relations.    

\section{Helium abundances of RR Lyrae: where the eagles dare}

Precise abundances of  primordial helium ($Y_p$) together with the abundances
of a few other light elements can provide robust constraints on the primordial 
nucleosynthesis, and in particular on the baryonic density of the Universe. 
Unfortunately, stellar spectra display strong photospheric helium absorption lines in the 
visual spectral range only at high effective temperatures ($T_e >$ 10,000 K). High-mass
stars are useless in constraining $Y_p$, since their material was already polluted
by previous stellar generations. Low-mass stars only attain hot effective temperatures
during central helium-burning phases. The typical evolutionary lifetime of these phases
--called Hot and Extreme Horizontal-Branch phases-- is of the order of 100 Myr. However,
these stellar structures cannot be adopted to constrain $Y_p$, since their surface
abundance  are affected by gravitational settling and/or by radiative levitation
\citep{Behr03,Moehler04}.

To overcome current observational limits, accurate estimates of  $Y_p$ have been
provided using He emission lines in HII region of blue compact galaxies.
Current estimates agree, within the errors, with the $Y_p$ abundance provided
by WMAP \citep{Larson11}. However, the agreement is not
solid because  $Y_p$ is used as a prior in the cosmological solutions. Moreover,
the spectroscopic abundances based on nebular lines might be affected by
systematic errors \citep{Bresolin09, Bresolin11}.

The observational scenario concerning the helium abundance has been recently
enriched by the possible occurrence of a variation in the helium abundance
of Globular Cluster (GC) stars. The presence
of He-enriched stars in GCs has been suggested to explain not only the presence
of multiple unevolved sequences, but also the presence of extended blue HB tails
\citep{Dantona08}. This working hypothesis relies on the
well established anti-correlations between the molecular band-strengths of
CN and CH \citep{Smith87,Kraft94} and between O--Na and Mg--Al  measured in 
evolved (RG, Horizontal Branch [HB]), 
and in unevolved (Main Sequence [MS]) stars of the GCs investigated 
so far with high-resolution spectra \citep{Pilachowski83, Gratton04}.
More recently, deep Hubble Space Telescope photometry disclosed the presence
of multiple stellar populations in several massive GCs. Together with $\omega$
Centauri \citep{Bedin04} multiple stellar sequences have been detected in
GCs covering a broad range of metal contents: NGC~2808 \citep{Piotto07},
M54 \citep{Siegel07} and NGC~1851 \citep{Calamida07,Milone08}. Some of these 
multiple sequences ($\omega$ Cen, NGC~2808, NGC~1851) might be explained 
either with a He-enhanced \citep{Norris04,Dantona08,Piotto07}, or with a 
CNO-enhanced \citep{Calamida07, Cassisi08} sub-population.
However, we still lack an empirical validation of the occurrence of He-enhanced
sub-population(s) in GCs.

To overcome the spectroscopic problems, \citet{Iben68} suggested 
to use the R-parameter --the number ratio between HB and RGB stars brighter 
than the luminosity level of the HB at the RR Lyrae instability strip (IS)-- 
to estimate the initial He abundance of cluster stars.\\ 
Two independent approaches to estimate the helium content in GCs were also suggested by 
\citet{Caputo83}. The $\Delta$-parameter,
--the difference in magnitude between the MS at (B-V)$_0$=0.7 and the luminosity
level of the HB at the RR Lyrae IS, and the A-parameter --the mass-to-luminosity
ratio of RR Lyrae stars. The pros and cons of the quoted parameters were
discussed in a thorough investigation by \citet{Sandquist00}. 
Theoretical and empirical limits affecting the precision of the R-parameter 
have been also discussed by \citet{Zoccali00}, 
\citet{Riello03} and \citet{Salaris04}.

The key feature of the quoted parameters is that they are directly or 
indirectly connected with the HB luminosity level. In spite of the improvements
in the photometric precision, in the sample of known cluster HB stars,
we still lack firm empirical methods to estimate the HB luminosity level
in GCs. This problem is partially due to substantial changes in the HB morphology
when moving from metal-poor (blue HB) to metal-rich (red HB) GCs. Moreover, we still
lack a robust diagnostic to constrain the actual off-ZAHB evolution of HB stars
(\citep{Ferraro99, Dicecco10, Cassisi11}.
To overcome several of the above drawbacks, \citet{troisi11} suggested a new method 
based on the difference in luminosity between the RGB bump and the main sequence 
benchmark at the same color of the bump. The new method shows several indisputable 
advantages, but also a strong dependence on the metal content.    

A new approach has been suggested by \citet{Dupree11} who
detected, in a few RG stars in $\omega$ Cen, the chromospheric
He~I line at 10830\AA. The He line was detected in more
metal-rich stars and it seems to be correlated with Al and Na, but no clear
correlation was found with Fe abundance. However, a more detailed non-LTE analysis
of the absolute abundance of He is required before firm conclusions can be drawn concerning 
the occurrence of a spread in helium content.

Interestingly enough, \citet{Preston09} discovered both He~I and
He~II emission and absorption lines in field RR Lyrae stars. The lines show up
during the rising branch of RR Lyrae stars and appear to be the consequence
of the shock propagation soon after the phases of minimum radius
\citep{Bono94b,Chadid96}.
The first detection of helium lines in low-mass variables dates back to
\citet{Wallerstein59}, who detected helium lines in Type II
Cepheids.  The key advantage of RR Lyrae stars, when compared with
detections of helium lines in hot and extreme HB stars (Behr 2003), is that
they have an extended convective envelope. Therefore, they are minimally
affected by gravitational settling and/or radiative levitation \citep{Michaud04}.
The drawback is that the measurement of the helium abundance requires hydrodynamical
atmosphere models accounting for time-dependent, convective transport, radiative
transfer equations, together with the formation and the propagation of sonic shocks.

Empirical evidence indicates that He absorption and emission lines in RR Lyrae
stars take place along the rising branch of the light curve. Plain
physical arguments suggest that they are triggered by the formation and the development
of strong shocks across the phases of maximum compression.  To further constrain the
physical mechanism(s) driving the occurrence of these interesting phenomena, new 
high-resolution, high signal-to-noise spectra are required to constrain the dependence 
of He lines on evolutionary and pulsation properties of RR Lyrae stars and eventually 
to estimate the helium abundance.

\section{Future remarks}

Recent theoretical and observational perspectives indicate that RR Lyrae stars are 
heading for a new golden age. This is due to the huge amount of field RR Lyrae stars that 
have already been collected by extended photometric surveys. A sample of $\sim$1,200 
field RR Lyrae were collected by the Northern Sky Variability Survey (NSVS, \citealt{woz04})  
and analyzed by \citet{Kinemuchi06}. They cover a distance of $\sim$7-9 Kpc in the solar 
neighborhood and reach a limit magnitude of V=15 mag. 
An even more ambitious project was realized by the All Sky Automated Survey (ASAS, 
\citealt{szcz09}), since they covered the southern
sky up to $\delta$=+28 ($\sim$ 75\% of their sky). The survey has a limiting magnitude of 
V$\sim$14 and they identified more than 1,450 RR Lyrae within 4 kpc in the solar neighborhood. 
Deep multiband photometric (ugriz) and spectroscopic data for almost 500 RR Lyrae have been collected 
by the Sloan Digital Sky Survey II (SDSS--II, \citealt{sesar10}, Sesar et al. these proceedings).
They have been able to investigate the spatial distribution of halo RR Lyrae stars up to 
galactocentric distances of 5-100 kpc. A sizable sample of field RR Lyrae (more than 2,000) 
was found by \citep{keller08} using V and R-band archival data collected by the 
Southern Edgeworth-Kuiper Belt Object (SEKBO) survey. This survey covers 1675 square degrees 
along the ecliptic to a mean depth of V=19.5, i.e. up heliocentric distances of $\sim$50 kpc.     
However, the real quantum jump concerning the photometric precision, the depth and the sampling 
of current RR Lyrae surveys, is given by the Optical Gravitational Lensing Experiment III 
survey (OGLE III, \citealt{pietr11}). They collected V and I-band data over a time interval 
of almost ten years,  toward the Galactic bulge and identified the richest sample ever 
collected of field RR Lyrae, i.e. more than 16,800 stars. 
The near future appears even more promising not only concerning the new optical (Pan--STARRS) 
and NIR (VISTA) photometric surveys, but also for the spectroscopic surveys (M2FS@Magellan, 
FLAMES@VLT). This means that together with pulsation properties, also the metallicity of a 
relevant fraction of field RR Lyrae will become available.  
    
It is noteworthy that to attack the open problems mentioned above requires a multiwavelength, 
a spectroscopic, and a theoretical approach \citep{Benko10,Marconi11}: a challenge that we have 
to deal with to further improve our knowledge of the detailed structure of the Galactic spheroid 
and to improve the accuracy of the RR Lyrae distance scale. The goal is not trivial: 
not only to further constrain the structure of the Galactic bulge and the interaction 
with the Galactic Bar, but also to trace the possible transition from thick to thin 
disc stars (Kinemuchi et al. 2006). 
Note that up to now, we still lack firm empirical evidence concerning the presence of old, 
low-mass stars --like RR Lyrae stars-- in the thin disc. There are a few field stars that are 
good candidates (TV Lib), but their intrinsic properties might be significantly different 
than typical RR Lyrae \citep{bono97}. The RR Lyrae can be even more relevant as beacons 
to trace the stellar streams in the Galactic halo (Marconi et al. 2006) and to constrain the 
outermost radial extent of the Galactic halo.  

This new spin will make a comeback to RR Lyrae and in this ongoing effort George Preston 
will always be a reference point for suggestions and new ideas.

\acknowledgments
One of us (G.B.) thanks G. Preston and A. McWilliam for the invitation and for the support to 
attend this exciting meeting. This project was partially supported by the PRIN-INAF 
(P.I.: R. Gratton).


\end{document}